\documentclass[11pt]{article}
\usepackage{amsfonts,amsmath,amssymb,epsfig,float,here,latexsym,setspace}
\usepackage[all]{xy}
\usepackage[square, comma]{natbib}
\usepackage[english]{babel}
\usepackage{hyperref}

\def\qed{\unskip\nobreak\hfill$\Box$\par\addvspace{\medskipamount}}

\newcommand{\be}{\begin{equation}}
\newcommand{\ee}{\end{equation}}
\newcommand{\bea}{\begin{eqnarray}}
\newcommand{\eea}{\end{eqnarray}}
\newcommand{\beas}{\begin{eqnarray*}}
\newcommand{\eeas}{\end{eqnarray*}}

\newtheorem{theorem}{Theorem}[section]

\newtheorem{remark}[theorem]{Remark}
\newtheorem{example}[theorem]{Example}
\newtheorem{examples}[theorem]{Examples}
\newtheorem{foo}[theorem]{Remarks}

\begin{document}
\title{\vskip -0.4cm
Risk Aversion in the Small and in the Large under Rank-Dependent Utility%: EU vs. DT
%\thanks{
%%We are very grateful to Johanna Etner and
%%Claudio Zoli for preliminary discussions on the topic.
%This research was funded in part by the Netherlands Organization for
%Scientific Research (Laeven) under grant NWO VIDI
%2009.}
%\vskip 1cm
}
\author{Louis R. Eeckhoudt\\
{\footnotesize IESEG School of Management}\\
{\footnotesize Catholic University of Lille}\\
{\footnotesize and CORE}\\
{\footnotesize {\tt Louis.Eeckhoudt@fucam.ac.be}}\\\and Roger J. A. Laeven\thanks{Corresponding author.
Mailing address: University of Amsterdam, Amsterdam School of Economics, PO Box 15867, 1001 NJ Amsterdam, The Netherlands.
Phone: +31 20 525 4219/4252.}\\
{\footnotesize Amsterdam School of Economics}\\
{\footnotesize University of Amsterdam, EURANDOM}\\
{\footnotesize and CentER}\\
{\footnotesize {\tt R.J.A.Laeven@uva.nl}}\\
[0.0cm]}
\date{This Version:
\today} \maketitle
\begin{abstract}
Under expected utility the local index of absolute risk aversion has played
a central role in many applications.
Besides, its link with the ``global'' concepts of the risk and probability premia has reinforced its attractiveness.
This paper shows that, with an appropriate approach, %appropriate comparison between a small risk and a constant situation
similar developments can be achieved in the framework of Yaari's dual theory
and, more generally, under rank-dependent utility.
%We revisit the risk and probability premia using the local risk aversion approach of Arrow \cite{A65,A71} and Pratt \cite{P64} in the expected utility model,
%introduce their appropriate counterparts in the dual and rank-dependent utility theories,
%and explicate the connections among these %six
%notions %, both analytically (
%by Taylor series approximations.
%%) and graphically.
%%We find that the risk (probability) premium under the rank-dependent utility model
%%simply equals the suitably scaled sum of the primal and dual risk (probability) premia.
%%Furthermore, we find
%%that the dual probability (risk) premium exactly mimics the primal risk (probability) premium.
%Our results %also %, appealing in their simplicity,
%%reveal
%illustrate
%the central role played by the ratio of the second and first derivatives of the
%probability weighting function %, $-\frac{h''}{h'}$,
%in dictating locally the risk and probability premia under the dual model
%and the interplay between this ratio and the well-known local index of absolute risk aversion of expected utility
%%,
%%$-\frac{U''}{U'}$,
%in dictating the premia under the rank-dependent utility model.
%%s, respectively.
%%and
%%Finally, they illustrate
%We also obtain the corresponding global properties in the dual and rank-dependent utility theories.
\noindent
\\[4mm]\noindent\textbf{Keywords:}
Risk Premium; Probability Premium; Expected Utility; Dual Theory; Rank-Dependent Utility;
%(Dual, Inverse) Stochastic Dominance;
Risk Aversion.
\\[4mm]\noindent\textbf{AMS 2010 Classification:} Primary: 91B06, 91B16, 91B30; Secondary: 60E15, 62P05.
%%60E15: inequalities, stochastic ordering
%%62P05: application to actuarial science or financial mathematics
%%91B06: Decision theory
%%91B16: Utility theory
%%91B30: Risk theory, insurance
\\[4mm]\noindent\textbf{OR/MS Classification:} Decision analysis: Risk.
\\[4mm]\noindent\textbf{JEL Classification:} D81, G10, G20.
\end{abstract}

\makeatletter
\makeatother
\maketitle

%\newpage

%\vskip 1cm

\onehalfspacing

\section{Introduction}

Under expected utility (EU) there exist various equivalent ways
to evaluate the degree of risk aversion of a decision maker (DM).
These measures were developed independently
a little more than 50 years ago by Arrow \cite{A65,A71} and Pratt \cite{P64}.\footnote{See also the early contribution (written in Italian) by B. de Finetti \cite{dF52}.}
Since then they have played a well-known and important role in the analysis of risky choices under EU.

The notion of risk aversion has also received attention outside EU.
For instance, in the dual theory (Yaari \cite{Y87}) or under rank-dependent utility (Quiggin \cite{Q82}),
Yaari \cite{Y86}, Chew, Karni and Safra \cite{CKS87}, Ro\"ell \cite{R87}, Chateauneuf, Cohen and Meilijson \cite{CCM04}, and Ryan \cite{R06}
have proposed various measures of risk aversion,
but their approach is rather different from the local risk aversion approach of Arrow and Pratt (henceforth AP) in the EU model.\footnote{One exception is
Yaari \cite{Y86} who proposes in a relatively little used paper the dual analog of the local index of absolute risk aversion.
However, his analysis focuses on \textit{global} aversion to mean-preserving spreads
via a concavification of the probability weighting function.}$^{,}$\footnote{As is well-known,
rank-dependent utility encompasses expected utility and the dual theory as special cases
and is at the basis of (cumulative) prospect theory (Tversky and Kahneman \cite{TK92}).
For measures of risk aversion under prospect theory, see Schmidt and Zank \cite{SZ08}.
}

The purpose of this paper is to show that the analysis ``\`a la AP'' can be developed both
in the dual theory (DT) and under rank-dependent utility (RDU).
This goes by defining the appropriate notion of a ``small risk'',
which is small in payoff under EU, but small in probability %mass
under DT,
and small in both under RDU.
It reveals that for small risks under RDU the risk premium and the probability premium
are simply the sums of the suitably scaled equivalent notions under EU and DT.
As a result, AP's approach turns out to be appropriate to measure the intensity of risk aversion also outside the EU model.

In this paper we refer to three measures of the degree of risk aversion:
the risk premium, the probability premium and the local index of absolute risk aversion.
Among these concepts, the probability premium was for a long time by far the least popular. % one.
However, because of its appealing nature, it starts now being discussed in the literature
(Jindapon \cite{J10}, Liu and Meyer \cite{LM13}, Liu and Neilson \cite{LN15}, and Eeckhoudt and Laeven \cite{EL15}).
Initially, the probability premium was discussed by Pratt \cite{P64}
for a binary zero mean and symmetric risk.
We will stick here to this approach,
even though it was very recently extended (Liu and Neilson \cite{LN15}).

Our analysis shows that the probability (risk) premium under DT,
obtained by defining the appropriate counterpart of the AP approach,
exactly mimics the risk (probability) premium under EU.
Our results also illustrate that the ratio of the
second and first derivatives of the probability weighting function of DT
plays a central role
in dictating locally the respective risk and probability premia,
and that the interplay between this ratio and the local index of absolute risk aversion of EU
dictates these premia under RDU.
We show furthermore that not only the local properties of AP's approach can be obtained under DT and RDU,
but that also the corresponding global properties are valid.

Our developments are made possible by appropriate comparisons between simple pairs of lotteries, just like under EU.
In particular, the AP local risk aversion approach under EU defines the risk and probability premia
by comparing a certain situation to a corresponding risky one with small payoff.
To define the risk and probability premia under DT and RDU,
we consider changes in situations that involve small probabilities for DT and joint small payoffs and small probabilities for RDU.
%compare a constant situation with small probability to an equal probability risky one.

Our paper is organized as follows.
In Section \ref{sec:EU} we recall the local risk aversion approach of AP and the definitions of the risk and probability premia under the EU model.
In Section \ref{sec:DT} we develop the appropriate counterparts under the DT model
and we provide the corresponding global properties in Section \ref{sec:CRADT}.
In Sections \ref{sec:RDU} and \ref{sec:CRARDU} we extend our development to the RDU model.
We conclude in Section \ref{sec:Con}.
%Figures and
Proofs are relegated to the Appendix.

\setcounter{equation}{0}

\section{Preliminaries under EU}\label{sec:EU}

\begin{figure}[H]
\begin{center}
\caption{Ex Ante and Ex Post Situation under EU
\newline\footnotesize This figure plots the initial situation and the situation after the addition of a small risk under EU.
%The parameters and values alongside the arrows represent probabilities
%while those at the end of the arrows represent payoffs.
}
\vskip 0.4 cm
\includegraphics[scale=1.22,angle=0]{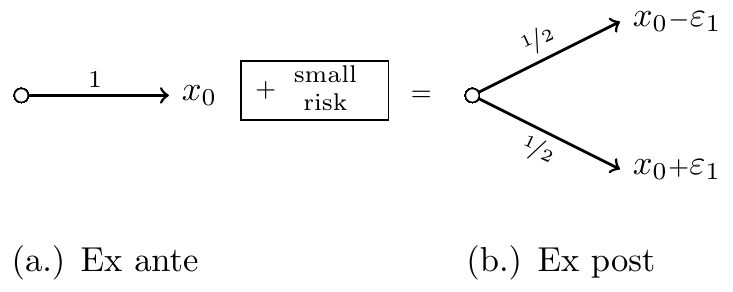}
\label{fig:ExAnteExPostEU}
\end{center}
\end{figure}

Consider an EU DM endowed with a certain initial wealth $x_{0}$
who takes a %small
binary symmetric risk that pays off $\pm \varepsilon_{1}$ each with probability $\frac{1}{2}$.
His ex ante and ex post situations can be represented by Figure \ref{fig:ExAnteExPostEU}. %, (b.).
(In all figures throughout this paper, the parameters and values alongside the arrows represent probabilities
and those at the end of the arrows represent payoffs.)

\vskip -0.5cm

\begin{figure}[H]
\begin{center}
\caption{Risk and Probability Premia under EU
\newline\footnotesize This figure visualizes the construction of the risk and probability premia under EU.
%The parameters and values alongside the arrows represent probabilities
%while those at the end of the arrows represent payoffs.
}
\vskip 0.4 cm
\includegraphics[scale=1.22,angle=0]{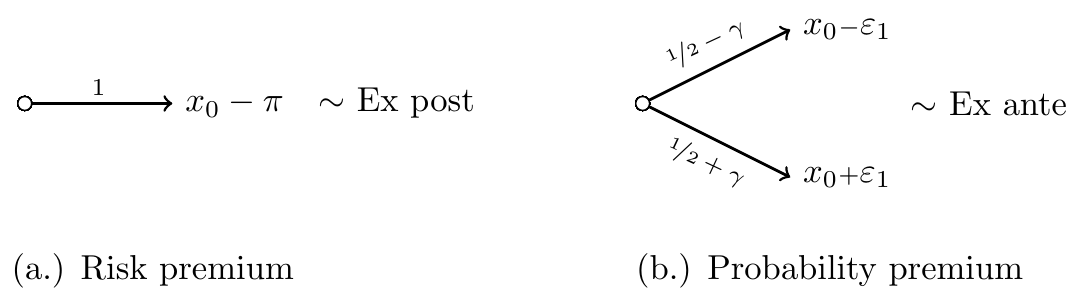}
\label{fig:RPPPEU}
\end{center}
\end{figure}

For this individual, the risk premium $\pi$ is defined such that $x_{0}-\pi$ with certainty is equivalent to the ex post situation.
The probability premium denoted by $\gamma$ is the increase in the probability of obtaining $x_{0}+\varepsilon_{1}$
that makes the DM indifferent between the then resulting situation and the initial situation in which $x_{0}$ occurs with certainty.
See the illustrations in Figure \ref{fig:RPPPEU}.
(In all figures, the symbol $\sim$ indicates indifference.)

\vskip -0.5cm

%\subsection{Risk premium under EU}
Formally, under EU, for a given initial wealth level $x_{0}$ and a given lottery payoff $\pm \varepsilon_{1}$ (with $\varepsilon_{1}>0$)
each occurring with probability $\frac{1}{2}$,
the risk premium $\pi$ is obtained as the solution to
\begin{equation}
U(x_{0}-\pi)=\frac{1}{2}U(x_{0}-\varepsilon_{1})+\frac{1}{2}U(x_{0}+\varepsilon_{1}),
\label{eq:rpEUeq}
\end{equation}
with $U$ the decision maker's (DM's) subjective utility function.
For ease of exposition $U$ is supposed to be twice continuously differentiable with $U'>0$.
As is well-known %since Pratt \cite{P64} and Arrow \cite{A65,A71},
$U''<0$ implies $\pi>0$.
%If we take $x_{0}=\frac{1}{2}$, $\varepsilon_{1}=\frac{1}{2}$, $U(0)=0$, and $U(1)=1$, then \eqref{eq:rpEUeq} simplifies to
%\begin{equation}
%U\left(\frac{1}{2}-\pi\right)=\frac{1}{2}.
%\end{equation}

As in the local risk aversion approach of Pratt \cite{P64} and Arrow \cite{A65,A71}, we approximate the solution to (\ref{eq:rpEUeq})
using Taylor series approximations.
Upon invoking a first order Taylor series expansion of $U(x_{0}-\pi)$ around $x_{0}$,
%$$U(x_{0}-\pi)=U(x_{0})-\pi U'(x_{0})+o(\pi),$$
and a second order Taylor series expansion of $U(x_{0}\pm \varepsilon_{1})$ around $x_{0}$,
%$$U(x_{0}\pm \varepsilon_{1})=U(x_{0})\pm \varepsilon_{1} U'(x_{0})+\frac{1}{2}\varepsilon_{1}^{2}U''(x_{0})+o(\varepsilon_{1}^2),$$
we obtain
%\begin{equation*}
%U(x_{0})-\pi U'(x_{0})=\frac{1}{2}\left(U(x_{0})-\varepsilon_{1} U'(x_{0})+\frac{1}{2}\varepsilon_{1}^{2}U''(x_{0})\right)
%+\frac{1}{2}\left(U(x_{0})+\varepsilon_{1} U'(x_{0})+\frac{1}{2}\varepsilon_{1}^{2}U''(x_{0})\right),
%\end{equation*}
%and, hence,
the well-known result
%$$-\pi U'(x_{0})=\frac{1}{2}\varepsilon_{1}^{2}U''(x_{0}),$$
%or
\begin{equation}
\pi=-\frac{1}{2}\varepsilon_{1}^{2}\frac{U''(x_{0})}{U'(x_{0})},
\label{eq:rpEU}
\end{equation}
omitting terms of order $o(\pi)$ and $o(\varepsilon_{1}^{2})$.

%Graphically, $\pi$ corresponds to the horizontal bar ... in Figure ... [***ADD FIGURE***]

%\subsection{Probability premium under EU}

Next, under EU, the probability premium $\gamma$ is obtained as the solution to
\begin{equation}
U(x_{0})=\left(\frac{1}{2}-\gamma\right)U(x_{0}-\varepsilon_{1})+\left(\frac{1}{2}+\gamma\right)U(x_{0}+\varepsilon_{1}).
\label{eq:ppEUeq}
\end{equation}
%Taking $x=\frac{1}{2}$, $\varepsilon_{1}=\frac{1}{2}$, $U(0)=0$, and $U(1)=1$, yields the simplification
%\begin{equation}
%\gamma=U(\frac{1}{2})-\frac{1}{2}.
%\end{equation}
Approximating the solution to (\ref{eq:ppEUeq}) by
invoking a second order Taylor series expansion of $U(x_{0}\pm \varepsilon_{1})$ around $x_{0}$
yields
%\begin{align*}
%U(x_{0})=&\left(\frac{1}{2}-\gamma\right)\left(U(x_{0})-\varepsilon_{1} U'(x_{0})+\frac{1}{2}\varepsilon_{1}^{2}U''(x_{0})\right)\\
%&+\left(\frac{1}{2}+\gamma\right)\left(U(x_{0})+\varepsilon_{1} U'(x_{0})+\frac{1}{2}\varepsilon_{1}^{2}U''(x_{0})\right),
%\end{align*}
%hence
%$$0=2\gamma\varepsilon_{1} U'(x_{0})+\frac{1}{2}\varepsilon_{1}^{2}U''(x_{0}),$$
%or
\begin{equation}
\gamma=-\frac{1}{4}\varepsilon_{1}\frac{U''(x_{0})}{U'(x_{0})},
\label{eq:ppEU}
\end{equation}
omitting terms of order $o(\varepsilon_{1}^{2})$.
Thus, upon comparing \eqref{eq:rpEU} and \eqref{eq:ppEU},
one finds that (the Taylor series approximations to) the primal risk and probability premia $\pi$ and $\gamma$ are linked through
\begin{equation}\label{eq:linkEU}
\pi=2\varepsilon_{1}\gamma.
\end{equation}
Under AP's local risk aversion approach, $\varepsilon_{1}$ is considered to be small
and $-\frac{U''}{U'}$ appearing in \eqref{eq:rpEU} and \eqref{eq:ppEU} is referred to as the local index of absolute risk aversion of EU.

%Graphically, $\gamma$ corresponds to the vertical bar ... in Figure ... [***ADD FIGURE***]

\setcounter{equation}{0}

\section{Risk and Probability Premia under DT}\label{sec:DT}

\begin{figure}[H]
\begin{center}
\caption{Small Risk\ \ \ \ \ \ \ \ \ \ \ \ \ \ \ \ \ \ \ \ \ \ \ \ \ \ \ \
\newline\footnotesize This figure plots the small binary symmetric risks
for EU and DT.
%The parameters and values alongside the arrows represent probabilities
%while those at the end of the arrows represent payoffs.
}
\vskip 0.4 cm
\includegraphics[scale=1.3,angle=0]{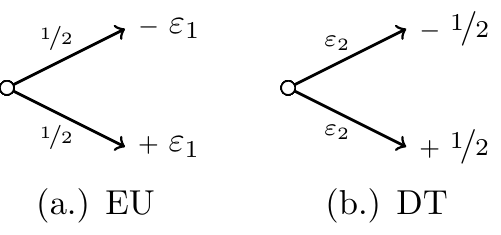}
\label{fig:smallEUDT}
\end{center}
\end{figure}

We now turn to DT.
%First, we note that under DT, a different definition of the risk premium is required.
Rather than considering a binary symmetric risk with small payoff $\pm\varepsilon_{1}$ each occurring with probability $\frac{1}{2}$ as under EU,
we now consider a binary symmetric risk with payoff $\pm\frac{1}{2}$ each occurring with small probability $\varepsilon_{2}$.
See the illustration in Figure \ref{fig:smallEUDT}.

\vskip -0.5cm

\begin{figure}[H]
\begin{center}
\caption{Ex Ante and Ex Post Situation under DT
\newline\footnotesize This figure plots the initial situation and the situation after attaching the small risk under DT.
%The parameters and values alongside the arrows represent probabilities
%while those at the end of the arrows represent payoffs.
}
\vskip 0.4 cm
\includegraphics[scale=1.22,angle=0]{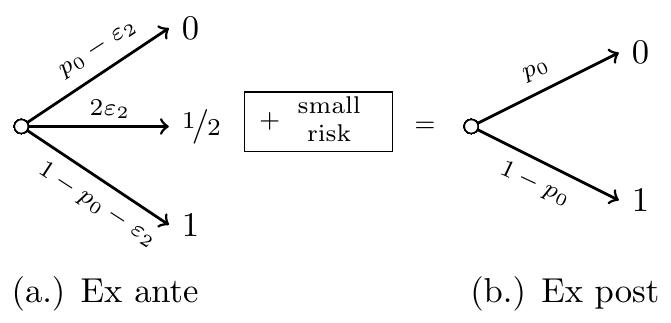}
\label{fig:ExAnteExPostDT}
\end{center}
\end{figure}

This binary symmetric risk with total probability mass of $2\varepsilon_{2}$ is added to an initial situation consisting of three states.
The initial situation equals $\frac{1}{2}$ (middle) with probability $2\varepsilon_{2}$
and divides the remaining probability mass between wealth levels 0 (low) and 1 (high), as follows:
$0$ with probability $p_{0}-\varepsilon_{2}$ and $1$ with probability $1-p_{0}-\varepsilon_{2}$
($0<\varepsilon_{2}\leq \{p_{0},1-p_{0}\}<1$).
The binary symmetric risk is attached to the middle state of the initial situation
such that, ex post, wealth level 0 occurs with probability $p_{0}$ and wealth level 1 occurs with probability $1-p_{0}$.
Attaching the small risk induces an equally likely change in payoff of $\pm \frac{1}{2}$
in a state with total probability mass of $2\varepsilon_{2}$.
See Figure \ref{fig:ExAnteExPostDT}.

\vskip -0.5cm

\begin{figure}[H]
\begin{center}
\caption{Risk and Probability Premia under DT
\newline\footnotesize This figure visualizes the construction of the risk and probability premia under DT.
%The parameters and values alongside the arrows represent probabilities
%while those at the end of the arrows represent payoffs.
}
\vskip 0.4 cm
\includegraphics[scale=1.22,angle=0]{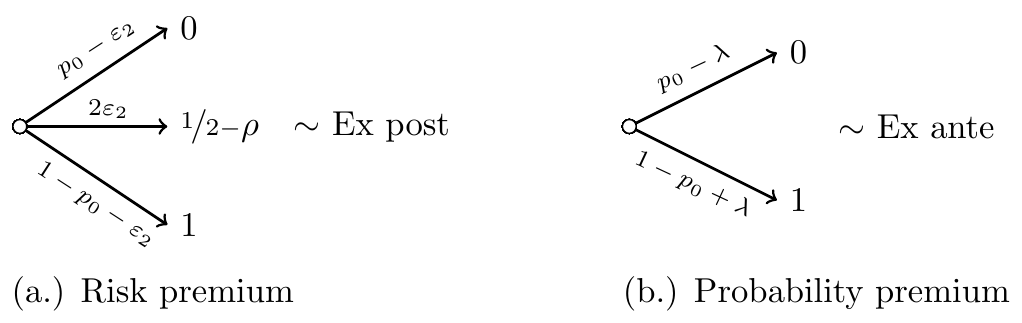}
\label{fig:RPPPDT}
\end{center}
\end{figure}

We then define the dual risk premium, now denoted by $\rho$ to distinguish it from that under EU,
such that indifference occurs between
%the addition of the binary symmetric risk to the middle state of the initial situation
the ex post situation
and a reduction of the wealth level (equal to the dual risk premium) in the middle state of the initial situation with total probability mass of $2\varepsilon_{2}$.\footnote{One may say that this constitutes a $2\varepsilon_{2}$-sure reduction in wealth, as opposed to a sure reduction in wealth as under EU.}
The dual probability premium $\lambda$ answers the question of by how much the probability %$\varepsilon_{2}$
of the unfavorable payoff %$-\frac{1}{2}$
should be reduced and shifted towards the favorable payoff %$\frac{1}{2}$
to restore equivalence with the initial situation.
See the illustrations in Figure \ref{fig:RPPPDT}.

\vskip -0.5cm

Formally, under DT, we define the risk premium $\rho$
as the solution to
\begin{align}
\left(h(p_{0}+\varepsilon_{2})-h(p_{0}-\varepsilon_{2})\right)&\left(\frac{1}{2}-\rho\right)\nonumber\\
=\left(h(p_{0})-h(p_{0}-\varepsilon_{2})\right)&\left(\frac{1}{2}-\frac{1}{2}\right)+\left(h(p_{0}+\varepsilon_{2})-h(p_{0})\right)\left(\frac{1}{2}+\frac{1}{2}\right),
\label{eq:rpDT0}
\end{align}
with $h$, mapping the unit interval onto itself with $h(0)=0$ and $h(1)=1$
and supposed to be twice continuously differentiable with $h'>0$,
the DM's subjective probability weighting (distortion) function.\footnote{\label{foot:DT}Under DT, an $n$-state risk with payoffs $0\leq x_{1}\leq\cdots\leq x_{n}$
and associated probabilities $p_{1},\cdots,p_{n}$
is evaluated according to
\begin{equation*}
\sum_{i=1}^{n}\left(h\left(\sum_{j=1}^{i}p_{j}\right)-h\left(\sum_{j=1}^{i-1}p_{j}\right)\right)x_{i},
\end{equation*}
with $\sum_{j=1}^{0}p_{j}=0$ by convention.
Equivalently, the $n$-state risk may be evaluated by distorting \textit{de}cumulative probabilities
rather than cumulative probabilities, as follows:
\begin{equation*}
\sum_{i=1}^{n}\left(\bar{h}\left(1-\sum_{j=1}^{i-1}p_{j}\right)-\bar{h}\left(1-\sum_{j=1}^{i}p_{j}\right)\right)x_{i},
\end{equation*}
with $\bar{h}(p)=1-h(1-p)$.
Note that $\bar{h}(0)=0$, $\bar{h}(1)=1$, $\bar{h}'>0$
and that $\bar{h}''>0$ is equivalent to $h''<0$.
The condition $h''<0$, referred to as ``strong risk aversion'',
means that the DM is averse to any mean preserving increase in risk (Chew, Karni and Safra \cite{CKS87} and Ro\"ell \cite{R87}).
Observe that the terms associated with the low and high wealth level states on the left- and right-hand sides of
\eqref{eq:rpDT0} partially cancel.
}
Observe that where the AP local risk aversion approach under EU defines the risk and probability premia
by comparing a certain situation to a corresponding risky one with small payoff,
\eqref{eq:rpDT0} defines the DT risk premium
by comparing a constant situation with small probability $2\varepsilon_{2}$ to an equal probability risky one.\footnote{The same will be true for the dual probability premium and for the risk and probability premia under RDU.}

Thus, we obtain the explicit solution
\begin{equation}
\rho=\frac{1}{2}\frac{\left(h(p_{0})-h(p_{0}-\varepsilon_{2})\right)-\left(h(p_{0}+\varepsilon_{2})-h(p_{0})\right)}
{h(p_{0}+\varepsilon_{2})-h(p_{0}-\varepsilon_{2})}.
\label{eq:rpDT}
\end{equation}
%If we take $\varepsilon_{2}=0$, this simplifies to
%\begin{equation}
%\rho=2h(\frac{1}{2})-1=\frac{(h(\frac{1}{2})-h(0))-(h(1)-h(\frac{1}{2}))}{h(1)-h(0)}.
%\end{equation}

Approximating the right-hand side of (\ref{eq:rpDT}),
by invoking a second order Taylor series expansion of $h(p_{0}\pm\varepsilon_{2})$ around $p_{0}$,
%$$h(p_{0}\pm\varepsilon_{2})=h(p_{0})\pm\varepsilon_{2}h'(p_{0})+\frac{1}{2}\varepsilon_{2}^2 h''(p_{0})+o(\varepsilon_{2}^{2}),$$
%for the numerator on the far RHS of (\ref{eq:rpDT})
%and a first order Taylor series expansion of $h(1)$ and $h(0)$ around $\frac{1}{2}-\varepsilon_{2}$
%for the denominator on the far RHS of (\ref{eq:rpDT})
yields
\begin{align}
\rho
%&=\frac{1}{2}\frac{-\varepsilon_{2}^2 h''(p_{0})}
%{2\varepsilon_{2}h'(p_{0})}\nonumber\\
&=-\frac{1}{4}\varepsilon_{2}\frac{h''(p_{0})}{h'(p_{0})},
\label{eq:rpDT2nd}
\end{align}
omitting terms of order $o(\varepsilon_{2}^{2})$.
Upon comparing \eqref{eq:ppEU} and \eqref{eq:rpDT2nd}
we find that the dual risk premium exactly mimics the primal probability premium.
We also find that the risk premium under DT is dictated locally by
the ratio of the second and first derivatives of the
probability weighting function, $-\frac{h''}{h'}$.

This dual local index of risk aversion first appeared in Yaari \cite{Y86},
in the context of a concavification of the DM's probability weighting function
to which a uniformly larger dual local index is shown to be equivalent,
when analyzing globally comparative risk aversion in the sense of
aversion to any mean-preserving spread.

%Graphically, $\rho$ corresponds to the vertical bar ... in Figure ... [***ADD FIGURE***]

%\subsection{Probability premium under DT}

Next, under DT, we define the probability premium $\lambda$ as the solution to
\begin{align*}
&\left(h(p_{0}+\varepsilon_{2})-h(p_{0}-\varepsilon_{2})\right)\frac{1}{2}\\
&\qquad=\left(h(p_{0}-\lambda)-h(p_{0}-\varepsilon_{2})\right)\left(\frac{1}{2}-\frac{1}{2}\right)+\left(h(p_{0}+\varepsilon_{2})-h(p_{0}-\lambda)\right)\left(\frac{1}{2}+\frac{1}{2}\right),
\end{align*}
which reduces to
\begin{equation}
0=\frac{1}{2}\left(h(p_{0}-\varepsilon_{2})-2h(p_{0}-\lambda)+h(p_{0}+\varepsilon_{2})\right).
\label{eq:ppDT}
\end{equation}
%If we take $\varepsilon_{2}=0$, this simplifies to
%\begin{equation}
%0=1-2h(\frac{1}{2}-\lambda)=-\frac{(h(\frac{1}{2}-\lambda)-h(0))-(h(1)-h(\frac{1}{2}-\lambda))}{h(1)-h(0)}.
%\end{equation}
From \eqref{eq:ppDT}, a first order Taylor series expansion of $h(p_{0}-\lambda)$ around $p_{0}$,
%$$h(p_{0}-\lambda)=h(p_{0})-\lambda h'(p_{0})+o(\lambda),$$
and a second order Taylor series expansion of $h(p_{0}\pm\varepsilon_{2})$ around $p_{0}$,
yield
%\begin{equation*}
$0=\frac{1}{2}\left(\varepsilon_{2}^{2} h''(p_{0})+2\lambda h'(p_{0})\right),
$
%\end{equation*}
hence
\begin{equation}
\lambda=-\frac{1}{2}\varepsilon_{2}^{2}\frac{h''(p_{0})}{h'(p_{0})},
\label{eq:ppDT2nd}
\end{equation}
omitting terms of order $o(\lambda)$ and $o(\varepsilon_{2}^{2})$.
Thus, by comparing \eqref{eq:rpDT2nd} and \eqref{eq:ppDT2nd},
we find that (the Taylor series approximations to) the dual risk premium $\rho$ and the dual probability premium $\lambda$ are linked through
\begin{equation}
\lambda=2\varepsilon_{2}\rho,
\end{equation}
and the local index $-\frac{h''}{h'}$ is central to both premia.
In particular, $\rho$ and $\lambda$ are both increasing with respect to this dual local index of absolute risk aversion
and with respect to the ``size'' of the risk represented by $\varepsilon_{2}$.
Observe also the similarity to the (reverse) link between the probability and risk premia under EU displayed in \eqref{eq:linkEU}.

%Graphically, $\lambda$ corresponds to the horizontal bar ... in Figure ... [***ADD FIGURE***]

\setcounter{equation}{0}

\section{Comparative Risk Aversion under DT}\label{sec:CRADT}

A concavification of an initial probability weighting function $h_{1}$
with $h_{1}''<0$
into $h_{2}$ via the transformation $h_{2}(p)=T(h_{1}(p))$,
where $T$ is a twice continuously differentiable function mapping the unit interval onto itself with $T'>0$, $T''<0$, $T(0)=0$ and $T(1)=1$,
yields
\begin{equation*}
-\frac{h_{2}''}{h_{2}'}>-\frac{h_{1}''}{h_{1}'}.
\end{equation*}
The analysis in the previous section has revealed that if the dual local index of absolute risk aversion
increases at a point $p_0$, then the corresponding risk and probability premia increase, too,
for ``small risks'', i.e., binary symmetric risks with payoff $\pm \frac{1}{2}$ each occurring with small probability $\varepsilon_{2}$;
see \eqref{eq:rpDT2nd} and \eqref{eq:ppDT2nd}.

In this section, we show that not just the local properties of the previous section hold,
but that also the corresponding global properties are valid.
That is, we obtain for the DT an analysis that parallels the one of Pratt \cite{P64}, pp. 127-129.
Specifically:
\begin{theorem}\label{th:craDT}
Let $h_{i}$, $\rho_{i}(p_{0},\varepsilon_{2})$ and $\lambda_{i}(p_{0},\varepsilon_{2})$ be
the probability weighting function, the risk premium \eqref{eq:rpDT} and the probability premium solving \eqref{eq:ppDT} for DT DM $i=1,2$.
Then the following conditions are equivalent:
\begin{itemize}
\item[(i)] $-\frac{h_{2}''(p)}{h_{2}'(p)}\geq -\frac{h_{1}''(p)}{h_{1}'(p)}$ $\qquad$ for all $p\in(0,1)$.
\item[(ii)] $\rho_{2}(p_{0},\varepsilon_{2})\geq \rho_{1}(p_{0},\varepsilon_{2})$ $\qquad$ for all $0<\varepsilon_{2}\leq\{p_{0},1-p_{0}\}<1$.
\item[(iii)] $\lambda_{2}(p_{0},\varepsilon_{2})\geq \lambda_{1}(p_{0},\varepsilon_{2})$ $\qquad$ for all $0<\varepsilon_{2}\leq\{p_{0},1-p_{0}\}<1$
\item[(iv)] $h_{2}(h_{1}^{-1}(t))$ is a concave function of $t$ $\qquad$ for all $t\in(0,1)$.
\item[(v)] $\frac{h_{2}(s)-h_{2}(r)}{h_{2}(q)-h_{2}(p)}\leq \frac{h_{1}(s)-h_{1}(r)}{h_{1}(q)-h_{1}(p)}$ $\qquad$ for all $0<p<q\leq r<s<1$.
\end{itemize}
\end{theorem}
The equivalence between (i) and (iv) can already be found in Yaari \cite{Y86}.

\setcounter{equation}{0}

\section{Risk and Probability Premia under RDU}\label{sec:RDU}

\begin{figure}[H]
\begin{center}
\caption{Small Risk
\newline\footnotesize This figure plots the small binary symmetric risks
for the three models for decision under risk: EU, DT and RDU.
%The parameters and values alongside the arrows represent probabilities
%while those at the end of the arrows represent payoffs.
}
\vskip 0.4 cm
\includegraphics[scale=1.3,angle=0]{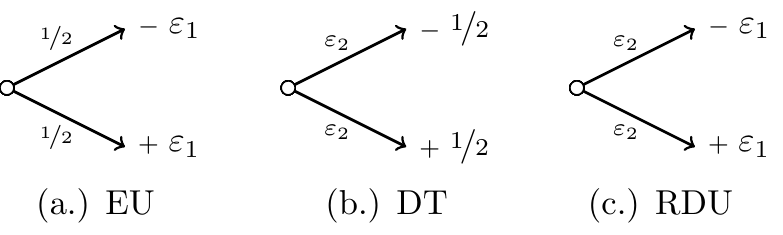}
\label{fig:smallEUDTRDU}
\end{center}
\end{figure}

We finally turn to RDU,
which encompasses both EU and DT as special cases when the utility function or the probability weighting are the identity function, respectively.
Under RDU we consider a binary symmetric risk that is small in both payoff ($\pm \varepsilon_{1}$)
and probability of occurrence ($\varepsilon_{2}$).
See the illustration in Figure \ref{fig:smallEUDTRDU}.

\vskip -0.5cm

\begin{figure}[H]
\begin{center}
\caption{Ex Ante and Ex Post Situation under RDU
\newline\footnotesize This figure plots the initial situation and the situation after attaching the small risk under RDU.
%The parameters and values alongside the arrows represent probabilities
%while those at the end of the arrows represent payoffs.
}
\vskip 0.4 cm
\includegraphics[scale=1.22,angle=0]{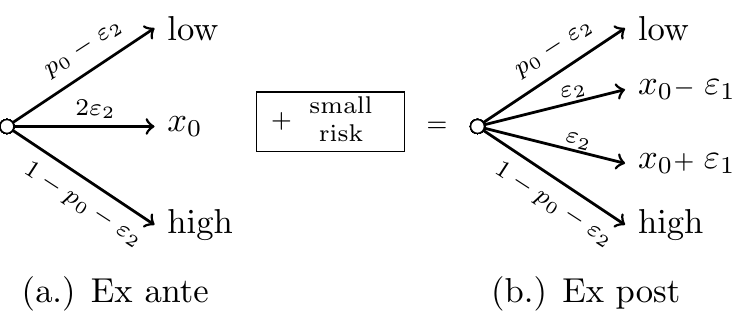}
\label{fig:ExAnteExPostRDU}
\end{center}
\end{figure}

This small risk is added to an initial situation consisting of three states:
a middle state at wealth level $x_{0}$ with total probability mass of occurrence equal to $2\varepsilon_{2}$,
and two states, one at a low wealth level and one at a high wealth level
(that can be left unspecified as long as they are at least $\varepsilon_{1}$ apart from $x_{0}$),
subdividing the remaining probability mass as $p_{0}-\varepsilon_{2}$ and $1-p_{0}-\varepsilon_{2}$ ($0<\varepsilon_{2}\leq\{p_{0},1-p_{0}\}<1$), respectively.
The small binary symmetric risk is attached to the middle state of the initial situation and
induces a payoff of $x_{0}\pm\varepsilon_{1}$ (with $\varepsilon_{1}>0$) each with probability $\varepsilon_{2}$,
all else equal.
See Figure \ref{fig:ExAnteExPostRDU}.

\vskip -0.5cm

\begin{figure}[H]
\begin{center}
\caption{Risk and Probability Premia under RDU
\newline\footnotesize This figure visualizes the construction of the risk and probability premia under RDU.
%The parameters and values alongside the arrows represent probabilities
%while those at the end of the arrows represent payoffs.
}
\vskip 0.4 cm
\includegraphics[scale=1.22,angle=0]{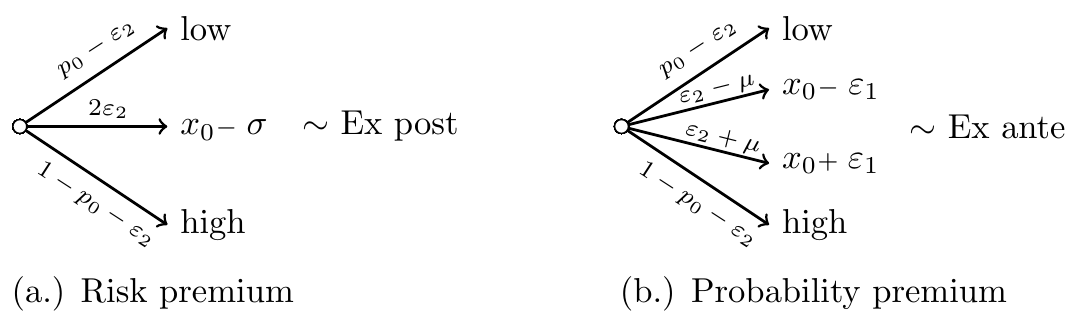}
\label{fig:RPPPRDU}
\end{center}
\end{figure}

The RDU risk premium $\sigma$ then achieves indifference between the ex post situation
(with the small risk added)
and a reduction in wealth (equal to the RDU risk premium)
in the middle state of the initial situation with probability of occurrence equal to $2\varepsilon_{2}$.
%Furthermore,
The RDU probability premium $\mu$ restores equivalence with the ex ante situation
by shifting probability mass from the unfavorable payoff to the favorable payoff.
See the illustrations in Figure \ref{fig:RPPPRDU}.

\vskip -0.5cm

Formally, under RDU, we define the risk premium $\sigma$ as the solution to\footnote{Under RDU,
an $n$-state risk with payoffs $0\leq x_{1}\leq\cdots\leq x_{n}$
and associated probabilities $p_{1},\cdots,p_{n}$
is evaluated according to
\begin{equation*}
\sum_{i=1}^{n}\left(h\left(\sum_{j=1}^{i}p_{j}\right)-h\left(\sum_{j=1}^{i-1}p_{j}\right)\right)U(x_{i});
\end{equation*}
cf. footnote \ref{foot:DT}.}
\begin{align}
&\left(h(p_{0}+\varepsilon_{2})-h(p_{0}-\varepsilon_{2})\right)U\left(x_{0}-\sigma\right)\nonumber\\
&\qquad=\left(h(p_{0})-h(p_{0}-\varepsilon_{2})\right)U(x_{0}-\varepsilon_{1})+\left(h(p_{0}+\varepsilon_{2})-h(p_{0})\right)U(x_{0}+\varepsilon_{1}).
\label{eq:rpRDUeq}
\end{align}
%If we take
%$x=\frac{1}{2}$, $\varepsilon_{1}=\frac{1}{2}$, $\varepsilon_{2}=0$, $U(0)=0$, and $U(1)=1$
%this simplifies to
%\begin{equation}
%U(\frac{1}{2}-\sigma)=1-h(\frac{1}{2}).
%\end{equation}
We approximate the solution to (\ref{eq:rpRDUeq}).
First, upon invoking a first order Taylor series expansion of $U(x_{0}-\sigma)$ around $x_{0}$,
and a second order Taylor series expansion of $U(x_{0}\pm \varepsilon_{1})$ around $x_{0}$,
we obtain
%\begin{align*}
%\left(h(p_{0}+\varepsilon_{2})-h(p_{0}-\varepsilon_{2})\right)&\left(U(x_{0})-\sigma U'(x_{0})\right)\\
%=\left(h(p_{0})-h(p_{0}-\varepsilon_{2})\right)&\left(U(x_{0})-\varepsilon_{1} U'(x_{0})+\frac{1}{2}\varepsilon_{1}^{2}U''(x_{0})\right)\\
%+&\left(h(p_{0}+\varepsilon_{2})-h(p_{0})\right)\left(U(x_{0})+\varepsilon_{1}U'(x_{0})+\frac{1}{2}\varepsilon_{1}^{2}U''(x_{0})\right),
%\end{align*}
%from which we solve
\begin{equation*}
\sigma=\varepsilon_{1}\frac{\left(h(p_{0})-h(p_{0}-\varepsilon_{2})\right)-\left(h(p_{0}+\varepsilon_{2})-h(p_{0})\right)}
{h(p_{0}+\varepsilon_{2})-h(p_{0}-\varepsilon_{2})}-\frac{1}{2}\varepsilon_{1}^{2}\frac{U''(x_{0})}{U'(x_{0})},
\end{equation*}
omitting terms of order $o(\sigma)$ and $o(\varepsilon_{1}^{2})$.
Next, recalling \eqref{eq:rpDT}
(and effectively taking a second order Taylor series expansion of $h(p_{0}\pm\varepsilon_{2})$ around $p_{0}$)
we obtain
\begin{equation}
\sigma=-\frac{1}{2}\varepsilon_{1}\varepsilon_{2}\frac{h''(p_{0})}{h'(p_{0})}
-\frac{1}{2}\varepsilon_{1}^{2}\frac{U''(x_{0})}{U'(x_{0})},
\label{eq:rpRDU}
\end{equation}
omitting terms of order $o(\varepsilon_{2}^{2})$.
Thus, we find that (the Taylor series approximations to) the RDU risk premium $\sigma$
is simply obtained as the suitably scaled sum of the primal risk premium $\pi$ and the dual risk premium $\rho$:
\begin{equation}
\sigma=\pi+2\varepsilon_{1}\rho.
\end{equation}
Observe also from \eqref{eq:rpRDU} the interplay between the local indexes $-\frac{h''}{h'}$ and $-\frac{U''}{U'}$
in dictating locally the RDU risk premium.

%\subsection{Probability premium under RDU}

Finally, we consider the probability premium under RDU,
denoted by $\mu$ and
defined as the solution to
\begin{align}
&\left(h(p_{0}+\varepsilon_{2})-h(p_{0}-\varepsilon_{2})\right)U(x_{0})\nonumber\\
&\qquad=\left(h(p_{0}-\mu)-h(p_{0}-\varepsilon_{2})\right)U(x_{0}-\varepsilon_{1})+\left(h(p_{0}+\varepsilon_{2})-h(p_{0}-\mu)\right)U(x_{0}+\varepsilon_{1}).
\label{eq:ppRDUeq}
\end{align}
%Taking
%$x=\frac{1}{2}$, $\varepsilon_{1}=\frac{1}{2}$, $\varepsilon_{2}=0$, $U(0)=0$, and $U(1)=1$
%this simplifies to
%\begin{equation}
%U(x)=1-h(\frac{1}{2}-\mu).
%\end{equation}
We approximate the solution to (\ref{eq:ppRDUeq})
by first invoking a second order Taylor series expansion of $U(x_{0}\pm \varepsilon_{1})$ around $x_{0}$
to obtain
\begin{align*}
2h(p_{0}-\mu)=h(p_{0}-\varepsilon_{2})+h(p_{0}+\varepsilon_{2})+\left(h(p_{0}+\varepsilon_{2})-h(p_{0}-\varepsilon_{2})\right)\frac{1}{2}\varepsilon_{1}\frac{U''(x_{0})}{U'(x_{0})},
\end{align*}
omitting terms of order $o(\varepsilon_{1}^{2})$.
Next, a first order Taylor series expansion of $h(p_{0}-\mu)$ around $p_{0}$,
and a second order Taylor series expansion of $h(p_{0}\pm\varepsilon_{2})$ around $p_{0}$,
yield
\begin{equation}
\mu=-\frac{1}{2}\varepsilon_{2}^{2}\frac{h''(p_{0})}{h'(p_{0})}-\frac{1}{2}\varepsilon_{1}\varepsilon_{2}\frac{U''(x_{0})}{U'(x_{0})},
\label{eq:ppRDU}
\end{equation}
omitting terms of order $o(\mu)$ and $o(\varepsilon_{2}^{2})$.
We thus find that the (Taylor series approximation to the) RDU probability premium, $\mu$,
is connected to the (Taylor series approximation to the) EU and DT probability premia, $\gamma$ and $\lambda$, through
\begin{equation}
\mu=2\varepsilon_{2}\gamma+\lambda.
\end{equation}
Observe also from \eqref{eq:rpRDU} and \eqref{eq:ppRDU} that the RDU risk and probability premia are linked through
\begin{equation}
\sigma=\frac{\varepsilon_{1}}{\varepsilon_{2}}\mu.
\end{equation}

\setcounter{equation}{0}

\section{Comparative Risk Aversion under RDU}\label{sec:CRARDU}

Just like under EU and DT, not only the local properties of the previous section are valid under RDU,
but also the corresponding global properties are true.
However, due to the simultaneous involvement of both the utility function and the probability weighting function,
the proof of the global properties under RDU is more complicated than that of the analogous properties under EU and DT.
Our proof is based on the total differential of the RDU evaluation,
and the sensitivities of the risk and probability premia
with respect to changes in outcomes and probabilities, respectively.

\begin{theorem}\label{th:craRDU}
Let $U_{i}$, $h_{i}$, $\sigma_{i}(x_{0},p_{0},\varepsilon_{1},\varepsilon_{2})$ and $\mu_{i}(x_{0},p_{0},\varepsilon_{1},\varepsilon_{2})$ be
the utility function, the probability weighting function, the risk premium solving \eqref{eq:rpRDUeq} and the probability premium solving \eqref{eq:ppRDUeq}
for RDU DM $i=1,2$.
Then the following conditions are equivalent:
\begin{itemize}
\item[(i)] $-\frac{U_{2}''(x)}{U_{2}'(x)}\geq -\frac{U_{1}''(x)}{U_{1}'(x)}$ and $-\frac{h_{2}''(p)}{h_{2}'(p)}\geq -\frac{h_{1}''(p)}{h_{1}'(p)}$
$\quad$ for all $x$ and all $p\in(0,1)$.
\item[(ii)] $\sigma_{2}(x_{0},p_{0},\varepsilon_{1},\varepsilon_{2})\geq \sigma_{1}(x_{0},p_{0},\varepsilon_{1},\varepsilon_{2})$
$\quad$ for all $x_{0}$, all $\varepsilon_{1}>0$, and all $0<\varepsilon_{2}\leq\{p_{0},1-p_{0}\}<1$.
\item[(iii)] $\mu_{2}(x_{0},p_{0},\varepsilon_{1},\varepsilon_{2})\geq \mu_{1}(x_{0},p_{0},\varepsilon_{1},\varepsilon_{2})$
$\quad$ for all $x_{0}$, all $\varepsilon_{1}>0$, and all $0<\varepsilon_{2}\leq\{p_{0},1-p_{0}\}<1$.
\item[(iv)] $U_{2}(U_{1}^{-1}(t))$ and $h_{2}(h_{1}^{-1}(u))$ are concave functions of $t$ and $u$ $\quad$ for all $t$ and all $u\in(0,1)$.
\item[(v)] $\frac{U_{2}(y)-U_{2}(x)}{U_{2}(w)-U_{2}(v)}\leq \frac{U_{1}(y)-U_{1}(x)}{U_{1}(w)-U_{1}(v)}$
and
$\frac{h_{2}(s)-h_{2}(r)}{h_{2}(q)-h_{2}(p)}\leq \frac{h_{1}(s)-h_{1}(r)}{h_{1}(q)-h_{1}(p)}$
$\quad$ for all $v<w\leq x<y$ and all $0<p<q\leq r<s<1$.
\end{itemize}
\end{theorem}

\section{Conclusion}\label{sec:Con}

We have extended the well-known Arrow-Pratt analysis of the risk and probability premia
under expected utility (EU) to the dual theory (DT) and rank-dependent utility (RDU) models
for decision under risk.
By adopting the appropriate notion of a ``small'' binary symmetric risk,
which is small in payoff under EU,
but small in probability %mass
under DT
and small in both under RDU,
we have developed Taylor series approximations to the DT and RDU risk and probability premia.
Our analysis has revealed that the development of the dual probability (risk) premium
exactly mimics that of the primal risk (probability) premium.
Furthermore, for small risks the RDU risk and probability premia appear to simply sum up the respective EU and DT premia,
upon suitable scaling.
Our analysis has also illustrated the central role of the probability weighting function's concavity index $-\frac{h''}{h'}$
and its interplay with the well-known utility function's concavity index $-\frac{U''}{U'}$
in dictating locally the DT and RDU risk and probability premia.
Finally, we have also obtained the corresponding global properties under DT and RDU.
%
%Specifically, denoting by
%\begin{alignat*}{2}
%&\pi:    \mathrm{the\ risk\ premium\ under\ EU}, \qquad&&\gamma:  \mathrm{the\ probability\ premium\ under\ EU};\\
%&\rho:   \mathrm{the\ risk\ premium\ under\ DT}, \qquad&&\lambda: \mathrm{the\ probability\ premium\ under\ DT};\\
%&\sigma: \mathrm{the\ risk\ premium\ under\ RDU},\qquad&&\mu:     \mathrm{the\ probability\ premium\ under\ RDU};
%\end{alignat*}
%we have shown that their local approximations satisfy
%\begin{align*}
%\pi=2\varepsilon_{1}\gamma,\qquad\lambda=2\varepsilon_{2}\rho,\qquad\sigma=\frac{\varepsilon_{1}}{\varepsilon_{2}}\mu;\\
%\sigma=\pi+2\varepsilon_{1}\rho,\qquad\mu=2\varepsilon_{2}\gamma+\lambda;
%\end{align*}
%and that their sizes are dictated by the primal and dual local indexes $-\frac{U''}{U'}$ and $-\frac{h''}{h'}$.

%In future work we intend to further develop the dual local index of absolute risk aversion.

In the EU model the local index of absolute risk aversion has been extremely useful
for the analysis of risky choices made by individuals or groups.
Since a similar approach seems to be appropriate also for alternative models of choice under risk,
this paper should open the way for their use in many applications such as portfolio composition, insurance coverage
or self-protection activities.

%\newpage

\appendix

%\section{Figures}

\setcounter{equation}{0}

%\newpage

\section{Proofs}

{\sc Proof of Theorem \ref{th:craDT}.}
Invoking the appropriate notion of
a binary symmetric risk that under DT has
%with
payoff $\pm\frac{1}{2}$ each occurring with probability $\varepsilon_{2}$ ($0<\varepsilon_{2}\leq\{p_{0},1-p_{0}\}<1$),
which is not necessarily small,
and upon replacing the equation for the EU risk premium, \eqref{eq:rpEUeq},
by that for the DT probability premium, \eqref{eq:ppDT},
and the equation for the EU probability premium, \eqref{eq:ppEUeq}, by the expression for the DT risk premium, \eqref{eq:rpDT},
the proof follows essentially the same arguments as the proof of Theorem 1 in Pratt \cite{P64}.
To save space we omit the details.
\qed

\noindent {\sc Proof of Theorem \ref{th:craRDU}.}
The equivalence of (i), (iv) and (v) follows trivially from the equivalence of (a), (d) and (e) in Theorem 1 in Pratt \cite{P64}
and the equivalence of (i), (iv) and (v) in Theorem \ref{th:craDT} above.

We will first prove that (the equivalent) (i), (iv) and (v) imply (ii) and (iii).
Reconsider \eqref{eq:rpRDUeq}.
Fix (a feasible) $\varepsilon_{2}>0$ (satisfying $0<\varepsilon_{2}\leq\{p_{0},1-p_{0}\}<1$).
Note that if we let $\varepsilon_{1}\rightarrow 0$ in \eqref{eq:rpRDUeq}, then $\sigma_{i}\rightarrow 0$.
Define
\begin{align*}
V_{i}(\sigma_{i},\varepsilon_{1})=&\left(h_{i}(p_{0}+\varepsilon_{2})-h_{i}(p_{0}-\varepsilon_{2})\right)U_{i}\left(x_{0}-\sigma_{i}\right)\nonumber\\
&-\left(\left(h_{i}(p_{0})-h_{i}(p_{0}-\varepsilon_{2})\right)U_{i}(x_{0}-\varepsilon_{1})
+\left(h_{i}(p_{0}+\varepsilon_{2})-h_{i}(p_{0})\right)U_{i}(x_{0}+\varepsilon_{1})\right).
\end{align*}
We compute the total differential
%\begin{equation*}
$\mathrm{d}V_{i}=\frac{\partial V_{i}}{\partial \sigma_{i}}\,\mathrm{d}\sigma_{i} + \frac{\partial V_{i}}{\partial \varepsilon_{1}}\,\mathrm{d}\varepsilon_{1}.
$
%\end{equation*}
It is given by
\begin{align*}
-&\left(h_{i}(p_{0}+\varepsilon_{2})-h_{i}(p_{0}-\varepsilon_{2})\right)U'_{i}\left(x_{0}-\sigma_{i}\right)\,\mathrm{d}\sigma_{i}\\
&+\left(\left(h_{i}(p_{0})-h_{i}(p_{0}-\varepsilon_{2})\right)U'_{i}(x_{0}-\varepsilon_{1})
-\left(h_{i}(p_{0}+\varepsilon_{2})-h_{i}(p_{0})\right)U'_{i}(x_{0}+\varepsilon_{1})\right)\,\mathrm{d}\varepsilon_{1}.
\end{align*}
Equating the total differential to zero yields
\begin{align}\label{eq:tdrp}
\frac{\mathrm{d}\sigma_{i}}{\mathrm{d}\varepsilon_{1}}=
\frac{h_{i}(p_{0})-h_{i}(p_{0}-\varepsilon_{2})}{h_{i}(p_{0}+\varepsilon_{2})-h_{i}(p_{0}-\varepsilon_{2})}\frac{U'_{i}(x_{0}-\varepsilon_{1})}{U'_{i}(x_{0}-\sigma_{i})}
-\frac{h_{i}(p_{0}+\varepsilon_{2})-h_{i}(p_{0})}{h_{i}(p_{0}+\varepsilon_{2})-h_{i}(p_{0}-\varepsilon_{2})}\frac{U'_{i}(x_{0}+\varepsilon_{1})}{U'_{i}(x_{0}-\sigma_{i})}.
\end{align}

From (i), as in Pratt \cite{P64}, Eqn. (20),
\begin{equation*}
\frac{U_{2}'(x)}{U_{2}'(w)}\leq \frac{U_{1}'(x)}{U_{1}'(w)},\qquad\mathrm{for}\ w<x,
%\end{equation*}
\qquad\mathrm{and}\qquad
%\begin{equation*}
\frac{U_{2}'(x)}{U_{2}'(y)}\geq \frac{U_{1}'(x)}{U_{1}'(y)},\qquad\mathrm{for}\ x<y.
\end{equation*}
Furthermore, from (v),
\begin{equation*}
\frac{U_{2}(y)-U_{2}(x)}{U_{2}(w)-U_{2}(v)}+\frac{U_{2}(w)-U_{2}(v)}{U_{2}(w)-U_{2}(v)}
\leq \frac{U_{1}(y)-U_{1}(x)}{U_{1}(w)-U_{1}(v)}+\frac{U_{1}(w)-U_{1}(v)}{U_{1}(w)-U_{1}(v)},\qquad\mathrm{for}\ v<w\leq x<y.
\end{equation*}
Taking $w=x$ yields
\begin{equation*}
\frac{U_{2}(y)-U_{2}(v)}{U_{2}(w)-U_{2}(v)}
\leq \frac{U_{1}(y)-U_{1}(v)}{U_{1}(w)-U_{1}(v)},\qquad\mathrm{for}\ v<w<y,
\end{equation*}
hence
\begin{equation*}
\frac{U_{2}(w)-U_{2}(v)}{U_{2}(y)-U_{2}(v)}
\geq \frac{U_{1}(w)-U_{1}(v)}{U_{1}(y)-U_{1}(v)},%\qquad\mathrm{for}\ v<w<y,
%\end{equation*}
\quad\mathrm{and\ also}\quad
%\begin{equation*}
\frac{U_{2}(y)-U_{2}(w)}{U_{2}(y)-U_{2}(v)}
\leq \frac{U_{1}(y)-U_{1}(w)}{U_{1}(y)-U_{1}(v)},
\end{equation*}
%\quad\mathrm{for}\ v<w<y.
for $v<w<y$.
In all inequalities in this paragraph, $U_{i}$ may be replaced by $h_{i}$, with $v,w,x$ and $y$ restricted to $(0,1)$.

Thus, from \eqref{eq:tdrp} and the inequalities above,
\begin{equation}\label{eq:tdrp2}
\frac{\mathrm{d}\sigma_{2}}{\mathrm{d}\varepsilon_{1}}\geq \frac{\mathrm{d}\sigma_{1}}{\mathrm{d}\varepsilon_{1}},
\end{equation}
hence (ii).

Next, reconsider \eqref{eq:ppRDUeq}.
Fix $\varepsilon_{1}>0$.
If we let $\varepsilon_{2}\rightarrow 0$ in \eqref{eq:ppRDUeq}, then $\mu_{i}\rightarrow 0$.
Define
\begin{align*}
W_{i}(\mu_{i},\varepsilon_{2})=
&\left(h_{i}(p_{0}+\varepsilon_{2})-h_{i}(p_{0}-\varepsilon_{2})\right)U_{i}(x_{0})\nonumber\\
&-\left(\left(h_{i}(p_{0}-\mu_{i})-h_{i}(p_{0}-\varepsilon_{2})\right)U_{i}(x_{0}-\varepsilon_{1})
+\left(h_{i}(p_{0}+\varepsilon_{2})-h_{i}(p_{0}-\mu_{i})\right)U_{i}(x_{0}+\varepsilon_{1})\right).
\end{align*}
We compute the total differential
%\begin{equation*}
$\mathrm{d}W_{i}=\frac{\partial W_{i}}{\partial \mu_{i}}\,\mathrm{d}\mu_{i} + \frac{\partial W_{i}}{\partial \varepsilon_{2}}\,\mathrm{d}\varepsilon_{2}.
$
%\end{equation*}
It is given by
\begin{align*}
-&h'_{i}(p_{0}-\mu_{i})\left(U_{i}(x_{0}+\varepsilon_{1})-U_{i}(x_{0}-\varepsilon_{1})\right)\,\mathrm{d}\mu_{i}\\
&+\left(h'_{i}(p_{0}-\varepsilon_{2})\left(U_{i}(x_{0})-U_{i}(x_{0}-\varepsilon_{1})\right)
-h'_{i}(p_{0}+\varepsilon_{2})\left(U_{i}(x_{0}+\varepsilon_{1})-U_{i}(x_{0})\right)\right)\,\mathrm{d}\varepsilon_{2}.
\end{align*}
Equating the total differential to zero yields
\begin{align}\label{eq:tdpp}
\frac{\mathrm{d}\mu_{i}}{\mathrm{d}\varepsilon_{2}}=
\frac{U_{i}(x_{0})-U_{i}(x_{0}-\varepsilon_{1})}{U_{i}(x_{0}+\varepsilon_{1})-U_{i}(x_{0}-\varepsilon_{1})}\frac{h'_{i}(p_{0}-\varepsilon_{2})}{h'_{i}(p_{0}-\mu_{i})}
-\frac{U_{i}(x_{0}+\varepsilon_{1})-U_{i}(x_{0})}{U_{i}(x_{0}+\varepsilon_{1})-U_{i}(x_{0}-\varepsilon_{1})}\frac{h'_{i}(p_{0}+\varepsilon_{2})}{h'_{i}(p_{0}-\mu_{i})}.
\end{align}

Invoking the inequalities associated with $\frac{U'_{i}(\cdot)}{U'_{i}(\circ)}$ and $\frac{U_{i}(\cdot)-U_{i}(\circ)}{U_{i}(\ast)-U_{i}(\star)}$ above,
where $U_{i}$ may be replaced by $h_{i}$ (upon restricting the corresponding domains to $(0,1)$),
we find from \eqref{eq:tdpp} that
\begin{equation}\label{eq:tdpp2}
\frac{\mathrm{d}\mu_{2}}{\mathrm{d}\varepsilon_{2}}\geq \frac{\mathrm{d}\mu_{1}}{\mathrm{d}\varepsilon_{2}},
\end{equation}
hence (iii).

We have now proved that (ii) and (iii) are implied by (the equivalent) (i), (iv) and (v).
We finally show that (ii) implies (i), and (iii) implies (i),
or rather that not (i) implies not (ii) and not (iii).
This goes by realizing that, by the arbitrariness of $x_{0}$, $p_{0}$, $\varepsilon_{1}>0$ and $\varepsilon_{2}$ with
$0<\varepsilon_{2}\leq\{p_{0},1-p_{0}\}<1$,
if (i) does not hold on some interval (of $x$ or $p$),
one can always find feasible $x_{0}$, $p_{0}$, $\varepsilon_{1}$ and $\varepsilon_{2}$, such that
\eqref{eq:tdrp2} and \eqref{eq:tdpp2}, hence (ii) and (iii), hold on some interval but with the inequality signs strict and flipped.
\qed

\textbf{Acknowledgements.}\
We are very grateful to Harris Schlesinger for discussions.
This research was funded in part by the Netherlands Organization for
Scientific Research (Laeven) under grant NWO VIDI 2009.
Research assistance of Andrei Lalu is gratefully acknowledged.

%\footnotesize

\begin{spacing}{0.0}

\end{spacing}

\end{document}